\documentclass[aps,twocolumn,groupedaddress,showpacs,floats,amssymb]{revtex4}
\usepackage[dvips]{graphicx}
\usepackage{bm}

\begin{document}

\noindent {\bf Comment on ``Comparative study of Higgs transition
in one-component and two-component lattice superconductor models"}\vspace{2mm}\\

In their recent paper \cite{MV}, Motrunich and Vishwanath report
results of Monte Carlo study of the Higgs transition in two
three-dimensional lattice realizations of the noncompact CP$^1$
model. The model---a gauge theory with two complex matter fields
with SU(2) invariance---is of great interest since it potentially
can yield a characteristic example of the so-called deconfined
critical point (DCP), provided it features a generic line of
second-order Higgs transitions. The authors claim an observation
of such a line in their simulations. In this comment, we question
the above conclusion, and show that the data of Ref.~\cite{MV}, if
properly processed, demonstrate the absence of scale invariance
and rather suggest a generic first-order line.

A significant progress in numerically distinguishing between
second-order and weak first-order transitions has been achieved
recently by utilizing the flowgram method \cite{flowgram}. In the
case when the existence of the tricritical point  separating the
first-order phase transition region (taking place at large enough
couplings) from the second-order phase transition region (taking
place at smaller couplings) is hard to reveal/rule out due to
system size limitations, it is crucially important to perform the
following data collapse analysis for flowgrams of quantities which
are supposed to be scale invariant at the second-order phase
transition. Namely, one rescales linear system size, $L \to
C(g)L$, where $C(g)$ is a certain smooth and monotonically
increasing function of the coupling constant $g=1/4K$ (usually, with
{\it a priori} known asymptotic behavior  at $g\ll 1$, which in
the present case is $C\propto g$ \cite{flowgram}). A collapse of
the rescaled flows in a given interval $g\in [0,\, g_{\rm coll}]$
implies that the tricritical point is either absent altogether, or
at least is located {\it outside} the interval $g\in [0,\, g_{\rm
coll}]$, so that the type of the transition within the interval
remains the same.

We note in passing that the flowgram method with data collapse
analysis has already allowed us to arrive at a definitive
conclusion about the generic first-order character of
the phase transition in {\it both} the U(1)$\times$U(1) DCP action
\cite{flowgram} and SU(2)-symmetric NCCP$^1$ model \cite{su2}.

The authors of Ref.~\cite{MV} do not perform the data collapse
study. They {\it interpret} the data presented in their Fig.~13 as
indicative of the tricritical point at $K\approx 0.2$ without
mentioning which features in particular are supporting their
conclusion. Here we perform the corresponding length-scale
renormalization analysis of the data digitized from Fig. 13 (lower panel) in
Ref.~\cite{MV}. We observe a convincing flow collapse with divergent flow
indicating the absence of scale invariance. This behavior is
inconsistent with the tricritical point at $K\approx0.2$. The scaling function $C(K)$ is chosen to be $C(0.2)=1$ and, then, two-parameter fit 
gave $C(K)=0.0740/K  + 0.00444*(\exp(1.00/K)-1)$. This
character of the flow is strikingly similar to the one found in
Refs.~\cite{flowgram,su2}.
Since the authors did not present flowgrams for the model NCCP II, we do not comment on the nature of the phases and transitions in it.

\begin{figure}[t]
\centerline{\includegraphics[scale=0.9, angle=0]{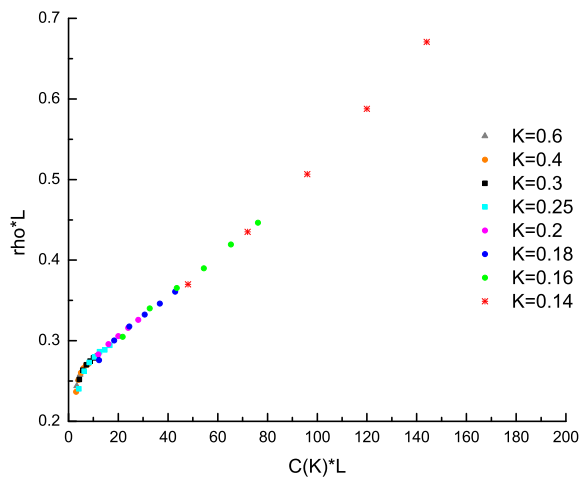}}
\caption{(Color online) Rescaled data from Fig.~13 in Ref.~\cite{MV}
demonstrating perfect collapse and diverging flow inconsistent with the scale invariance.} \label{Fig1}
\end{figure}

This work was supported by NSF under Grants Nos.
PHY-0653183 and PHY-0653135.\vspace{2mm}\\
{\small \noindent  Anatoly Kuklov$^1$, Munehisa Matsumoto$^{2}$, Nikolay Prokof'ev$^{3,4}$,
Boris Svistunov$^{4}$, and Matthias Troyer$^{3}$\vspace{2mm}\\
$^1$Department of Engineering \& Physics, The College of
Staten Island, CUNY, Staten Island, NY
10314.\\
$^2$Department of Physics, University of  
California, Davis, CA 95616\\
$^3$Institut f\"{u}r Theoretische Physik, ETH Z\"urich, CH-8093 Z\"urich. \\
$^4$Department of Physics,  University of Massachusetts,
Amherst, MA 01003.\vspace{2mm}




\begin{thebibliography}{99}

\bibitem{MV} O.I.~Motrunich and A.~Vishwanath, arXiv:0805.1494.

\bibitem{flowgram}  A.~Kuklov, N.~Prokof'ev, B.~Svistunov, and M.~Troyer, Ann. of Phys. {\bf 321}, 1602 (2006).

\bibitem{su2} A.~Kuklov, M.~Matsumoto, N.~Prokof'ev, B.~Svistunov, and
M.~Troyer, in preparation. Preliminary results have been presented
by A.~Kuklov at the Nordita {\it Quantum Fluids} workshop
(Stockholm, August 15 - September 30, 2007) http://www.nordita.org/~qf2007/kuklov.pdf ; and APS March Meeting
2008 (New Orleans), abstract: S12.00006.

\end{thebibliography}
\end{document}